\documentclass[a4paper,fleqn]{cas-dc}

\usepackage[english]{babel}
\usepackage[utf8]{inputenc}
\usepackage[T1]{fontenc}
\usepackage{microtype}
\usepackage{amsmath}
\usepackage{amssymb}
\usepackage{graphicx}
\usepackage{natbib}
\usepackage{nicefrac}
\usepackage{caption}
\usepackage{subcaption}
\usepackage{booktabs}
\usepackage{arydshln}
\usepackage{xcolor}
\usepackage[normalem]{ulem}
\usepackage{hyperref}
\usepackage[nameinlink, capitalise]{cleveref}

\makeatletter
\renewcommand{\fnum@figure}{Fig. \thefigure}
\makeatother

\def\tsc#1{\csdef{#1}{\textsc{\lowercase{#1}}\xspace}}
\tsc{WGM}
\tsc{QE}

\begin{document}
\let\WriteBookmarks\relax
\def\floatpagepagefraction{1}
\def\textpagefraction{.001}

\title [mode = title]{A simulation study on spatial and time resolution for a cost-effective positron emission particle tracking system}  

\author[1]{Josephine Oppotsch}[type=editor, 
							  style=chinese, 
							  orcid=0000-0003-2577-2069]
\cormark[1]
\cortext[1]{~corresponding author. E-mail address:}
\ead{joppotsch@ep1.rub.de}
\author[1]{Matthias Steinke}[style=chinese, orcid=0000-0002-4153-5488]
\author[1]{Miriam Fritsch}[style=chinese, orcid=0000-0002-6463-8295]
\author[1]{Fritz-Herbert Heinsius}[style=chinese, orcid=0000-0002-9545-5117]
\author[1]{Thomas Held}[style=chinese, orcid=0000-0001-8581-4238]
\author[2]{Nikoline Hilse}[style=chinese, orcid=0000-0002-8455-5251]
\author[2]{Viktor Scherer}[style=chinese]
\author[1]{Ulrich Wiedner}[style=chinese, orcid=0000-0002-9002-6583]

\affiliation[1]{organization={Ruhr-Universität Bochum},
				addressline={Institut für Experimentalphysik I}, 
				city={44801 Bochum},
				country={Germany}}
\affiliation[2]{organization={Ruhr-Universität Bochum},
				addressline={Lehrstuhl für Energieanlagen und Energieprozesstechnik}, 
				city={44801 Bochum},
				country={Germany}}

\begin{abstract}
This work is the second part of a simulation study investigating the processing of densely packed and moving granular assemblies by positron emission particle tracking (PEPT). Since medical PET scanners commonly used for PEPT are very expensive, a PET-like detector system based on cost-effective organic plastic scintillator bars is being developed and tested for its capabilities. In this context, the spatial resolution of a resting positron source, a source moving on a freely designed model path, and a particle motion given by a DEM (Discrete Element Method) simulation is studied using Monte Carlo simulations and the software toolkit Geant4. This not only extended the simulation and reconstruction to a moving source but also significantly improved the spatial resolution compared to previous work by adding oversampling and iteration to the reconstruction algorithm. Furthermore, in the case of a source following a trajectory developed from DEM simulations, a very good resolution of about 1\,mm in all three directions and an average three-dimensional deviation between simulated and reconstructed events of 2.3\,mm could be determined. Thus, the resolution for realistic particle motion within the generic grate system (which is the test rig for further experimental studies) is well below the smallest particle size. The simulation of the dependence of the reconstruction accuracy on tracer particle location revealed a nearly constant efficiency within the entire detector system, which demonstrates that boundary effects can be neglected.
\end{abstract}

\begin{keywords}
\sep Positron emission particle tracking 
\sep Monte Carlo method
\sep Time-of-flight
\sep Plastic scintillators
\sep Silicon photomultipliers (SiPMs)
\sep \rule{4.35cm}{0.3pt}
\sep License:
\sep $\copyright$ 2023. This manuscript version is made available under the CC-BY-NC-ND 4.0 license
\sep https://creativecommons.org/licenses/by-nc-nd/4.0/ 
\end{keywords}

\maketitle

\section{Introduction}\label{introduction}
\noindent Environmental protection and effective resource management play an essential role in today's world. One way to make an effective contribution is to improve large-scale industrial applications. Significant for a large part of these industrial applications is the processing of densely packed and moving granular material. However, due to the dense particle packing, it is quite difficult to get comprehensive bulk internal information about the particle motion in such a system experimentally. Therefore, a cost-effective positron emission tomography (PET) like detector system is under construction that allows for the study of particle movement via positron emission particle tracking (PEPT). In PEPT, a positron-emitting radioisotope is placed within the vessel under investigation. Due to the short mean free path of the emitted positrons (0.53\,mm on average, \cite{positronRangeNa-22}), they rapidly annihilate with electrons of the surrounding matter to a pair of 511\,keV back-to-back gamma rays. The point of origin of the two gamma rays can then be calculated since it must lie on a straight line connecting the two detection points (antiparallel emission). A comprehensive overview of the underlying physics, already existing PEPT systems, tracking algorithms, and radiolabeling techniques, as well as simulations and applications of PEPT systems, can be found in the work of \citet{PEPTReview}. In the scope of this work, the position of the tracer particle will be determined by calculating the time-of-flight (TOF) differences of the back-to-back gamma-ray pairs produced by positron-electron annihilation.

Since its development at the University of Birmingham \citep{PEPTDevelopment1,PEPTDevelopment2}, PEPT has been widely used in the context of process engineering. However, most detector systems used for these studies (such as the ADAC at the University of Birmingham) are based on expensive crystal scintillators \citep{PEPTReview}. Therefore, this work provides details of a new, less expensive detector system consisting of plastic scintillators instead of crystals.

Plastic scintillators are considerably cheaper to manufacture compared to scintillation crystals. However, they are composed of elements of low atomic numbers, so there is virtually no photoelectric cross-section for gamma rays of typical energies. Instead, gamma rays are detected mainly by Compton scattering; thus, no gamma energy measurement is performed. This does not affect the measurements in this study because the positions are determined solely by the time-of-flight differences of the back-to-back gamma-ray pairs.

Other major differences between the proposed detector system and conventional PET scanners/PEPT detectors are the arrangement of the scintillators and the design of the readout electronics. While most commonly used PEPT detector systems consist of a pixel design (such as dual-headed planar scanners consisting of two large area scintillation crystals coupled to an array of multiple photodetectors, ring scanners consisting of multiple, smaller detector crystals, or modular systems consisting of individual detector blocks (these blocks can also be sourced from conventional ex-clinical scanners), which are grouped into units called ``buckets'' \citep{PEPTReview}), the proposed detector system consists of long (1\,m) scintillator bars that are read out with photodetectors only at their respective ends. Therefore, the design of the proposed detector system saves a lot of photodetectors and readout electronics, which in turn reduces the overall cost.  The ``lack'' of granularity is then compensated for by calculating the impact points of the gamma rays onto the scintillator bars in the same way as the position of the tracer particle via TOF. 

To get a rough idea of the cost of a PEPT detector, take a look at the following exemplary calculation: The cheapest way of building such a big detector system (as the proposed one) would be to reuse and convert a former ex-clinical PET scanner. The first clinical PET scanner being able to perform TOF was the Philips Gemini TF. It is based on 28,336 4\,mm~$\times$~4\,mm~$\times$~22\,mm lutetium-yttrium oxyorthosilicate (LYSO) crystals, which are distributed on 28 arrays consisting of 33~$\times$~44 crystals each and read out by a hexagonal array of 420 photomultiplier tubes with a diameter of 39\,mm (PIXELAR Technology; \cite{PhillipsGeminiTF, surti2007performance}). This scanner can be purchased pre-owned for \$175,000 - \$225,000 \citep{GeminiTFkostenGebraucht}. However, to be able to investigate the generic grate system, which is the selected vessel for the upcoming experimental investigations (see \cref{Sec:ggs} for more details), the scanner has to be converted and expanded (2 detector walls of 1\,m~$\times$~0.5\,m should be enough to get sufficient resolution with the crystal scintillators). This means that at least 2.2 Gemini TF scanners are needed to cover the area of both walls with scintillators, which equates to at least \$385,000. Since only complete scanners can be purchased, three scanners would have to be bought, resulting in a total of at least \$525,000. In contrast, the budget of the proposed detector system is 200,000 euros (about \$219,495). Hence, switching to long plastic scintillator bars that are read out only at their respective ends saves at least \$305,500. However, this price only includes the three prefabricated scanners. Therefore, with unknown risks and cost estimates, they have to be disassembled and then reassembled to meet the requirements. As already described in the previous paragraph, the channel number of a commercially available system is significantly higher than for the system proposed in this work. Of course, this also means that significantly more readout electronics would be required if ex-clinical scanners were used instead of the proposed system, increasing the cost even further. However, since the basic cost of a commercial system is already significantly higher, this price difference was not calculated any further.

Being the second part of a simulation study on particle tracking in densely packed and moving granular assemblies with a PET-like detector system \citep{oppotsch2023simulation}, the focus this time is on improving the simulation and reconstruction and extending it to moving particles. Moreover, the envisaged detector system was revised to better fit the generic grate system. In the course of this, the number of scintillator bars has been reduced from a total of 192 to 88, which in turn led to a corresponding reduction of the readout electronics. This not only increases the solid angle coverage (by removing some of the scintillator bars while maintaining the same bar height, the walls can be moved closer to the vessel so that the opening above and below the generic grate system becomes smaller) but also leads to a significant reduction in the overall cost of the detector system. However, it is still possible to expand the detector to a larger size if necessary.

In order to improve accuracy in localizing the position of the tracer particle, the reconstruction algorithm itself has been extended to include the features of iteration and oversampling (for more details, see \cref{Sec:conditionsAndTracking}). In addition, the homogeneity of the reconstruction was investigated. This is important because the recorded point cloud of the reconstructed positions can be truncated if the tracer particle gets too close to the scintillators. However, when the center of gravity is formed from such a truncated point cloud, it is systematically shifted with respect to the actual position (this is referred to as boundary effects within this work). Since the generic grate system is small compared to the detector system, homogeneous reconstruction within the considered volume can be ensured so that the reconstruction must not be adapted to boundary effects. 

The detection efficiency is dependent on the threshold since only the Compton spectrum is available due to the use of plastic scintillators. For the simulations, a threshold of 30\,keV is used (see \cref{Sec:conditionsAndTracking}). Averaged over the detector volume, this results in an efficiency of about 1.4\,\% \citep{oppotsch2023simulation}.

Concerning the structure of this paper, \cref{setup} gives a brief overview of the underlying setup. 
This includes the detector geometry (\cref{Sec:detector}), vessel (\cref{Sec:ggs}), and granular assembly (\cref{Sec:spheres}) under investigation, as well as the source (\cref{Sec:source}), boundary conditions, and tracking method (\cref{Sec:conditionsAndTracking}) used. 
Following the setup, the results are presented in \cref{Sec:results}. 
To conclude, \cref{Sec:outlook} gives a brief summary and outlook on the upcoming work.

\hfill\\[-0.52cm]
\section{Setup} \label{setup}
\noindent For the simulation studies, the intended detector system, including the vessel and granular assembly, has been implemented in software. This has been done by using the simulation toolkit Geant4 (Geometry and Tracking 4). Based on the Monte Carlo method, it is an open-source, object-oriented development environment for simulating the interaction and passage of elementary particles through matter. A deeper insight into this toolkit is provided by \cite{Geant4SimToolkit} and \cite{Geant4DevelopmentsAndApplications, Geant4RecentDevelopments}.

Within the following work, a system consisting of the targeted detector and a vessel containing the granular assembly is considered. A description of the detector system, the vessel, and the granular assembly, as well as the positron source, boundary conditions, and reconstruction method used, is given in the following subsections.

\subsection{Detector system} \label{Sec:detector}
\begin{figure*}
\begin{subfigure}[b]{0.41\textwidth}
 	\centering
    \includegraphics[width=\textwidth]{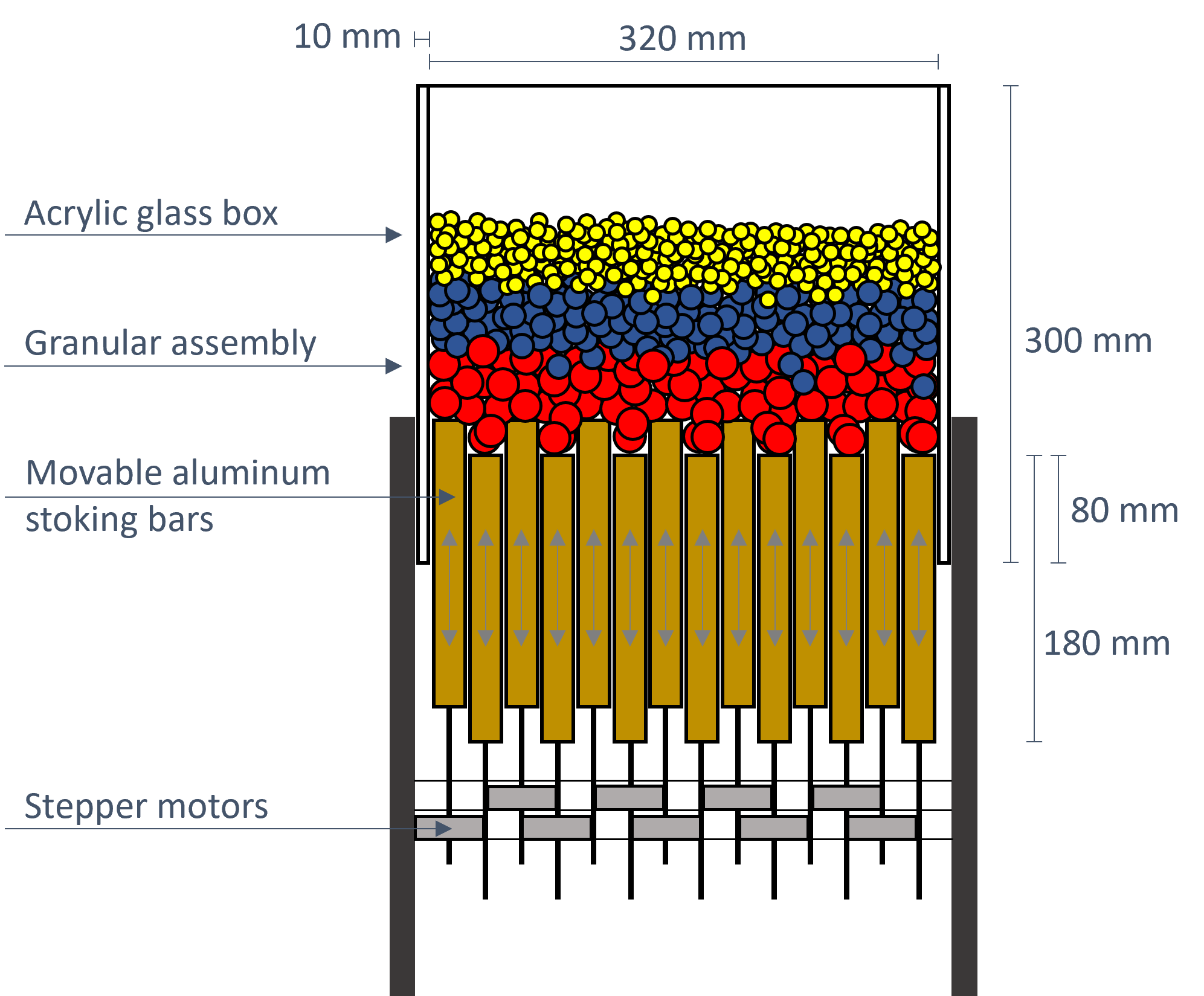}
    \caption{Schematic of the generic grate system.}
    \label{Fig.sketch}
\end{subfigure}
\hfill
\begin{subfigure}[b]{0.2\textwidth}
	\includegraphics[width=\textwidth]{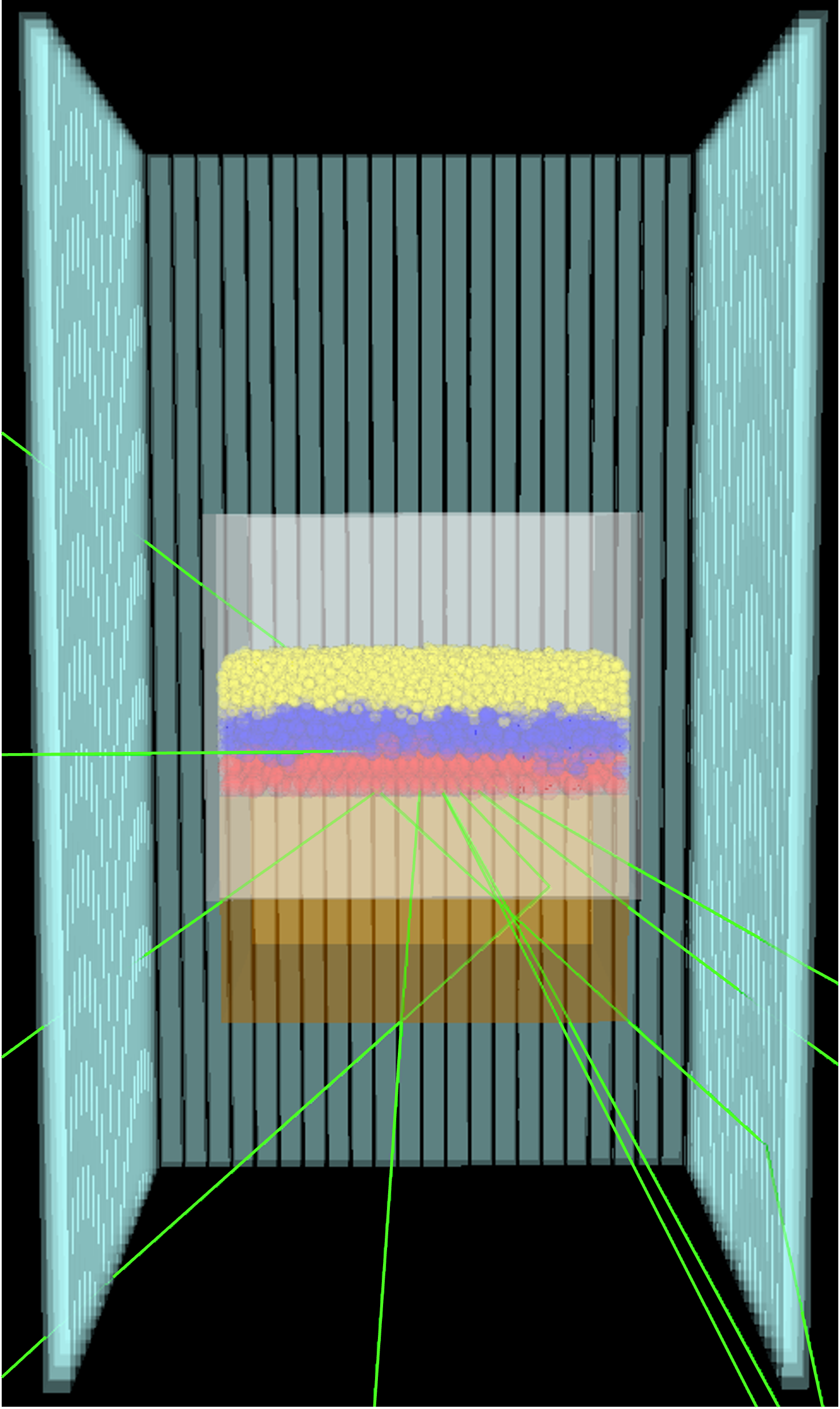}
    \caption{Simulated design.}
    \label{Fig.detectorWithGGS}
\end{subfigure}
\hfill
\begin{subfigure}[b]{0.337\textwidth}
	\includegraphics[width=\textwidth]{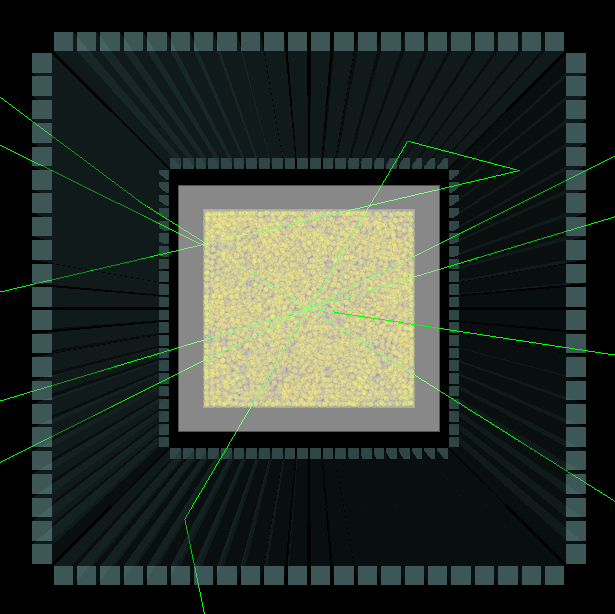}
    \caption{Top view of the simulated design.}
    \label{Fig.topView}
\end{subfigure}
\caption{Illustrations of the generic grate system and the detector system consisting of the scintillator bars. For simplicity, the photodetectors and readout electronics are not included. (a) shows a schematic representation of the grate system, and (b) and (c) simulated images of the grate system enclosed by the scintillator bars (turquoise). The green lines represent the gamma-ray traces.}
\label{Fig.grateSystem}
\end{figure*}

\begin{figure*}
 	\centering
    \includegraphics[width=0.98\textwidth]{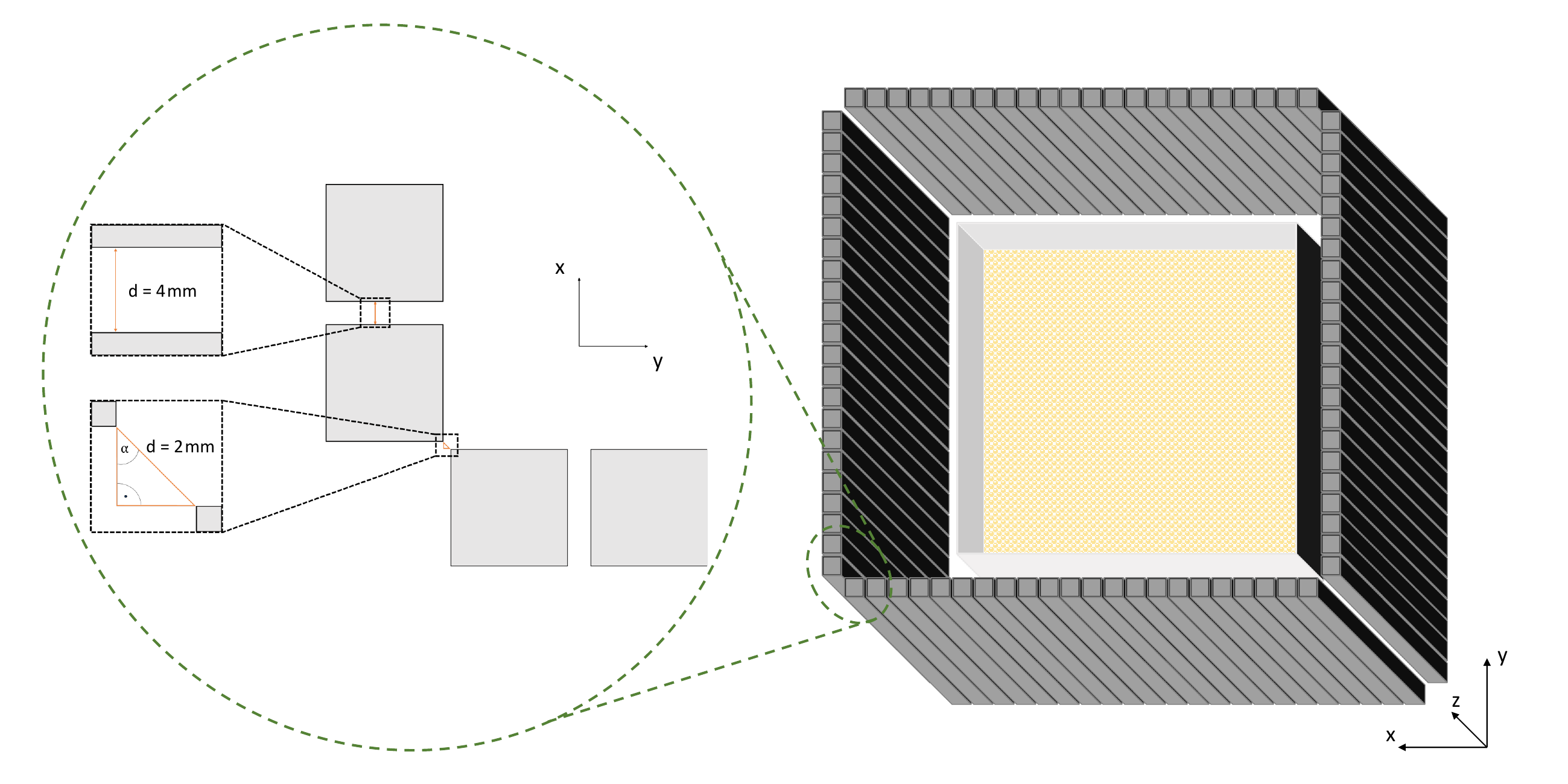}
    \caption{Sketch of the detector system with zoom to the front left corner to illustrate the distances between the individual detector walls and scintillator bars. The photodetectors and readout electronics are not included in the drawing.}
    \label{Fig.distanceBetweenWalls}
\end{figure*}

\noindent The simulated detector system consists of four walls. Each wall has a height of 1\,m and a length of 0.524\,m and is composed of 22 vertically arranged, cost-effective plastic scintillator bars with a length of 1\,m and a width and depth of 20\,mm each (see \cref{Fig.detectorWithGGS} and \ref{Fig.topView}). The individual bars are placed with a pitch of 24\,mm, as the bars will be wrapped in reflective foil (to maximize the light yield) and black foil (to shield from ambient light) in the upcoming experiments. This corresponds to a spacing of 4\,mm between the individual scintillator bars. The (diagonal) distance between the adjacent detector walls is set to 2\,mm (see also \cref{Fig.distanceBetweenWalls}).

Thus, all in all, the detector system (\cref{Fig.detectorWithGGS} and \ref{Fig.topView}) comprises 88 scintillator bars surrounding the vessel under investigation from four sides. Hence, both the vessel and the origin of the coordinate system (0\,|\,0\,|\,0) are exactly in the center of the detector system. 

\subsection{Generic grate system}\label{Sec:ggs}
\noindent The vessel chosen for the simulations is a generic grate system (GGS), which corresponds to the one used in the upcoming experiments. It is inspired by industrial grate firing systems, which are used, for example, for the combustion of wood chips or the incineration of municipal waste. However, since the main interest of this work is in the analysis of mixing and segregation, the grate system is simplified so that particle feeding and discharge are neglected (i.e., the grate system is operating in batch mode). Mixing and segregation of the granular assembly are caused by vertically movable aluminum stoking bars. Four acrylic glass walls are placed around the stoking bars, spanning a volume of 320\,mm $\times$ 300\,mm $\times$ 300\,mm and containing the granular assembly. A schematic and the simulated geometry of the generic grate system are shown in \cref{Fig.grateSystem}. For more information and results on mixing and segregation within the GGS, see the work of \cite{PaperNikoline}.

\hfill\\[-0.52cm]
\subsection{Granular assembly} \label{Sec:spheres}
\noindent In total, 6,064 polyoxymethylene (POM) spheres of three different sizes (4,357 yellow spheres with a diameter of 10\,mm, 1,207 blue spheres with a diameter of 15\,mm, and 500 red spheres with a diameter of 20\,mm) are considered as granular assembly. Polyoxymethylene (CH$_2$O)$_\text{n}$ is an engineering thermoplastic with a mass density of 1.4\,g/cm$^3$ predefined in the materials list of Geant4. The position of each sphere is taken from a DEM simulation (for more information concerning the DEM simulation, see \ref{App:A}). All spheres are kept static, even within the moving scenarios. This is done to keep the code simple and to reduce computational effort. To still be able to simulate particle motion, the decay location within the volume is moved instead of the spheres.

Since the spheres are at rest, the motion of the spheres and their collisions are not simulated. The influence of the spheres and their interspaces on the gamma rays, on the other hand, is simulated but with resting spheres. Even under consideration of sphere movement, the effect would most likely have canceled out at the end of the trajectory of the gamma rays to the detector by averaging.

\hfill\\[-0.92cm]
\subsection{Positron source} \label{Sec:source}
\hfill\\[-0.52cm]
\noindent The positron-emitting radioisotope used is $^{22}$Na since it is a sufficiently long-living source with a half-life of 2.6 years. It decays into the first excited state of $^{22}$Ne by emitting a positron with an average energy of 216\,keV or by electron capture, followed by a direct relaxation to the ground state by emitting a gamma ray with an energy of 1275\,keV. Only a small fraction of 0.055\% decays directly by emitting a positron with an average energy of 835\,keV into the ground state of $^{22}$Ne. The most abundant (90\,\%) decay is:
\begin{align}
_{11}^{22}\text{Na} \rightarrow \, _{10}^{22}\text{Ne} + \gamma\text{(1275\,keV)} + e^+ + \nu_{\text{e}}.
\end{align}
After 0.53\,mm on average or 2.28\,mm maximum (determined for water, \cite{positronRangeNa-22}), the positrons directly annihilate with electrons of the surrounding matter into 511\,keV gamma-ray pairs. This range is negligible compared to the radius of the tracer spheres, so the positrons will not leave the tracer sphere volume.

In the course of this work, only positrons are simulated and not the whole $^{22}$Na source. Since the positrons cannot leave the tracer spheres (maximum distance 2.28\,mm $\ll$ 10\,mm), their energy was chosen to be very small so that a point source can be assumed. Even though the radioactively labeled volume of a tracer will never be point-like in reality, this assumption was made because previous simulations (in which the motion of positrons was accounted for by a uniform distribution over a range of energies from 0 to 545\,keV) showed that the travel distance of the positrons did not strongly affect the spatial resolution. In addition to the small mean free path length of the positrons, another explanation is that the positrons move in 4$\pi$ so that the `blur' cancels out while averaging.

Each positron is assigned a random direction, which is determined by the Monte Carlo method. Moreover, generating them in any sphere within the detector system is possible. As they pass through the surrounding matter, they are subject to scattering, ionization, bremsstrahlung, and annihilation, which is accounted for by Geant4. 

For the reconstruction, it is roughly assumed that the gamma rays produced by positron-electron annihilation reach the scintillator bars in straight lines. However, if they have to pass through a lot of surrounding matter, they are subject to physics processes such as the photoelectric effect and Compton scattering. Thus, these processes are also considered by Geant4.

At this point, it should be noted that the spatial resolution will be somewhat worse for real measurements due to the assumptions made in this section (stationary positrons instead of the whole $^{22}$Na). However, the input parameters currently used in the simulations (such as the time resolution and threshold, \cref{Sec:conditionsAndTracking}) have a much greater impact on the resolution than these approximations. Unfortunately, these parameters can only be adjusted after the detector is completed.

\hfill\\[-1.5cm]
\subsection{Boundary conditions and tracking method} \label{Sec:conditionsAndTracking}
\hfill\\[-0.52cm]
\noindent For the sake of simplicity, the required photodetection and readout electronics of the detector system are implemented only in terms of their component properties. This is done by defining some boundary conditions (listed in \cref{Tab:boundaryConditions}), which are based on the component properties measured under real conditions in an experimental test setup. 

\begin{table}
\centering
\caption{Boundary conditions.} 
	\begin{tabular}{p{4cm} p{3cm}} 
	\toprule[1pt]
	\textbf{Property} 		  & \qquad\quad\textbf{Value} \\
	\hline	\\[-8pt]
	Time resolution           & \qquad\quad 460\,ps \\
	Pulse intensity threshold & \qquad\quad 30\,keV \\
	Activity 				  & \qquad\quad 10\,MBq~~ \\[1.5pt]
	\toprule[1pt]
	\end{tabular}
\label{Tab:boundaryConditions}
\end{table}

In the course of this work, a cut on the signal amplitude corresponding to an energy of 30 keV for a scintillation point close to the photodetector and an activity of 10\,MBq were chosen. The time resolution was set to 460\,ps (determined from the measured standard deviation).

\begin{figure*}
\begin{subfigure}[b]{0.49\textwidth}
    \includegraphics[width=\textwidth]{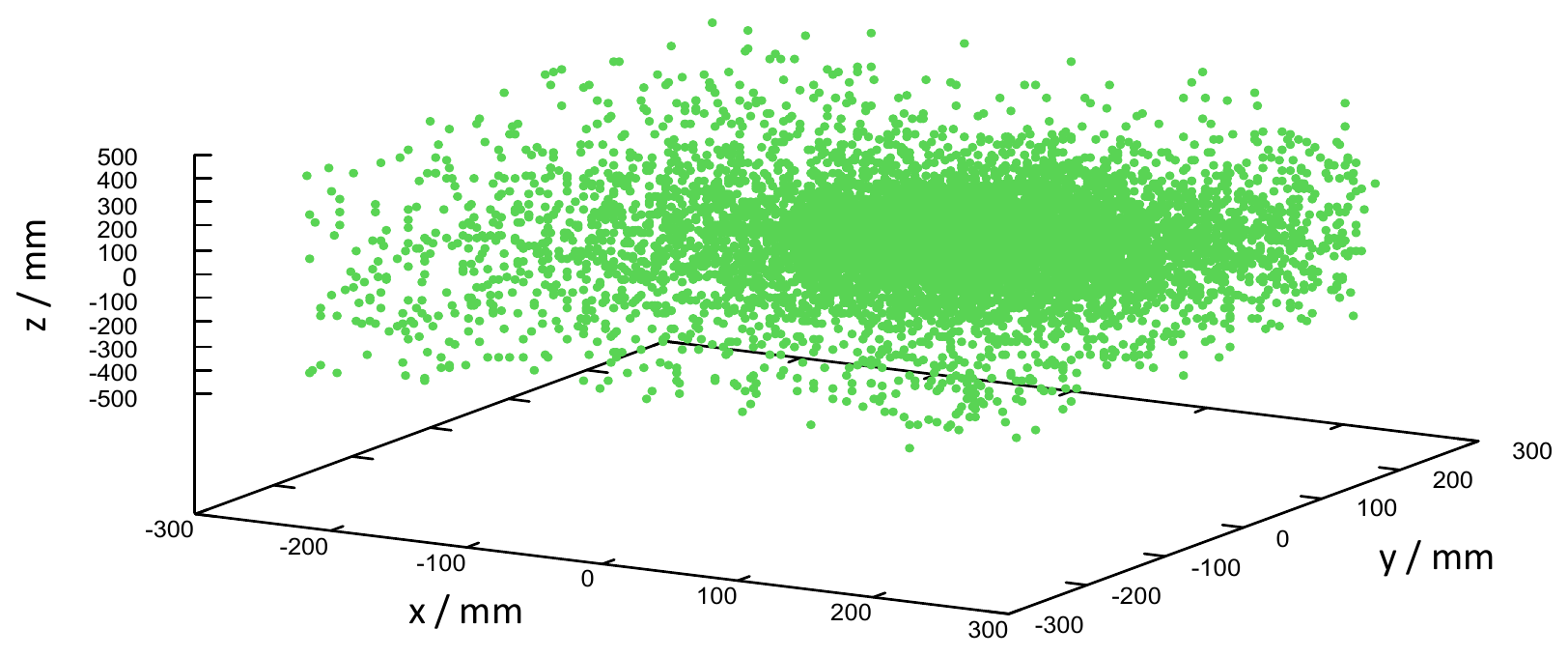}
    \caption{3D histogram of the reconstructed events $\vec{P}_{\text{\scriptsize{$S,k$}}}$.}
    \label{Fig.stationary3D}
\end{subfigure} 
\begin{subfigure}[b]{0.49\textwidth}
	\includegraphics[width=\textwidth]{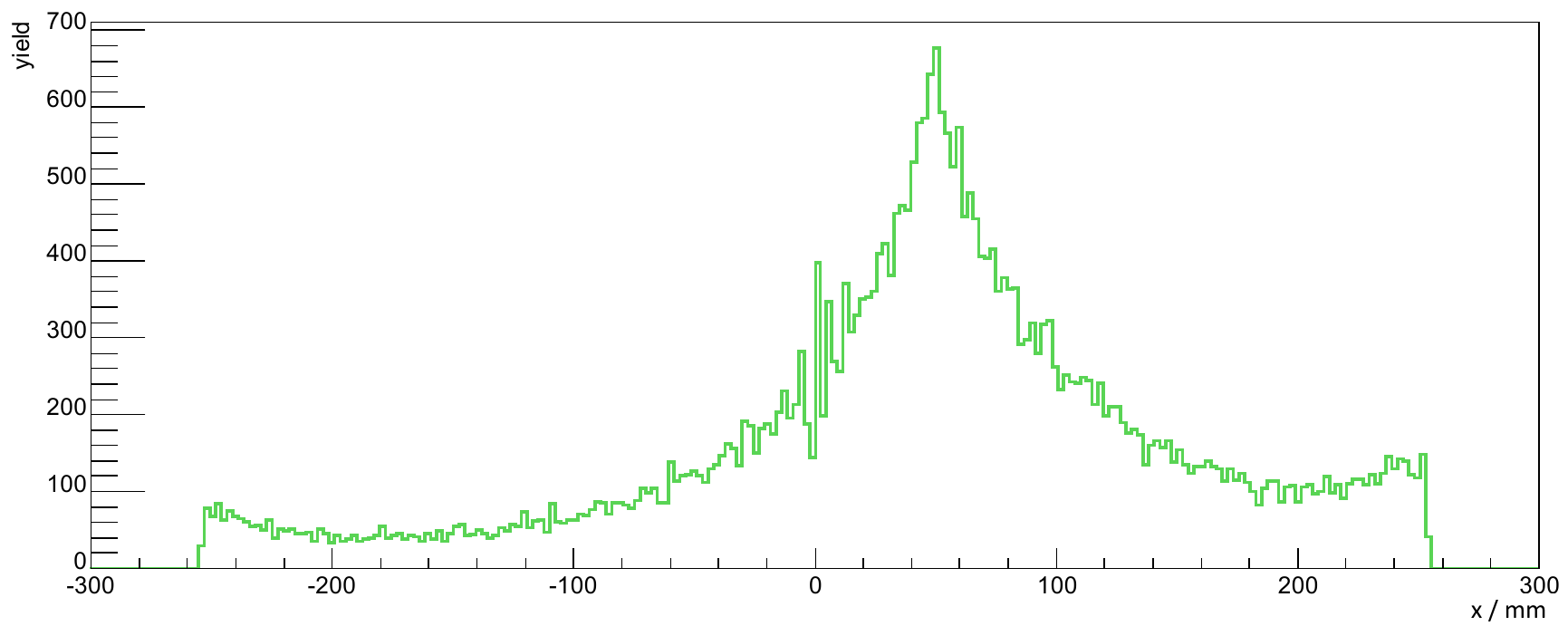}
    \caption{1D projection of the reconstructed events $\vec{P}_{\text{\scriptsize{$S,k$}}}$ in x-direction.}
    \label{Fig.stationaryX}
\end{subfigure} 
\caption{Spatial distribution of the reconstructed events $\vec{P}_{\text{\scriptsize{$S,k$}}}$ produced by a stationary source placed at (50\,|\,50\,|\,-40)\,cm. (a) shows all reconstructed events in a 3D perspective and (b) a projection of a complete buffer, meaning 36,000 events, in the x-direction.}
\label{Fig.stationarySource}
\end{figure*}

Analog to the work done in \cite{oppotsch2023simulation}, both the position of the gamma-ray hits onto the scintillator bars and the position of the tracer sphere are reconstructed by calculating the time-of-flight differences of the back-to-back gamma-ray pairs (TOF). Since every gamma-ray pair provides its individual position information of the tracer sphere $\vec{P}_{\text{\scriptsize{$S,k$}}}$, the most probable position $\vec{P}$ must be determined in another step. This is done by taking the mean of all collected position information in a reasonably selected time interval (the larger the buffer, the better the resolution. However, if the source is moving, the buffer size should be adjusted to the velocity of the source so that meaningful averaging can take place and the individual source positions are not offset against each other):
\begin{align}
\vec{P} = \frac{1}{n}\sum\limits_{k=1}^{n}\vec{P}_{\text{\scriptsize{$S,k$}}}
\label{eq:mean}
\end{align}
with $n$ as the number of individual tracer sphere positions. However, as briefly indicated in \cref{Sec:source}, not only unscattered but also scattered gamma rays are included in the reconstruction. This becomes evident in \cref{Fig.stationarySource}, which shows the individually calculated tracer sphere positions $\vec{P}_{\text{\scriptsize{$S,k$}}}$ for all gamma-ray pairs produced by a stationary tracer sphere in 1 second measurement time. Looking at the 3D histogram of \cref{Fig.stationarySource}, each green point corresponds to a reconstructed event $\vec{P}_{\text{\scriptsize{$S,k$}}}$ of the stationary source. It can be clearly seen that the entire GGS volume is filled with reconstructed events, forming a denser core around the true tracer sphere position. All events distributed around the core are positions reconstructed by scattered gamma-ray pairs, resulting in a non-constant background that affects the determination of the true source position and thus degrades its resolution. Now, if all reconstructed positions $\vec{P}_{\text{\scriptsize{$S,k$}}}$, including the background, are averaged, the reconstructed tracer sphere position $\vec{P}$ would differ from the actual position of the source. For example, if the mean is taken over the entire detector while the source is located in the corner of the detector, there will be a lower background in the direction of the adjacent walls, but a higher background in the direction of the center of the detector volume, so that the reconstructed position of the tracer sphere is shifted toward the center of the detector system.

To avoid this offset, the position components are filled into separate one-dimensional histograms from which the bin (one for x, y, and z each) with the most entries is selected. Then, merged into a three-dimensional spatial coordinate, this position information represents the centroid of the core. Finally, a cuboid is placed around this centroid, and all individual tracer sphere positions that fall into the cuboid are averaged (as in \cref{eq:mean}), resulting in a new centroid. With this new centroid, the procedure just described starts again (defining a new cuboid around the newly found centroid and averaging all the position information lying inside the cuboid).

In addition, the determination of the source position of a moving sphere can be further improved by calculating a moving average in the reconstruction. This is done by taking the first average over 36,000 events and the next every 18,000 new events by combining the 18,000 new data points with the last 18,000 events from the previous averaging. Thus, 36,000 events are required for one image point, but two times as many image points are generated with a moving average than with simple averaging. The number of 36,000 events corresponds to a buffer (in PEPT literature, also referred to as a sample) size adjusted for the particle motion given by a DEM simulation. Choosing a buffer size that is too small will degrade resolution, whereas choosing a buffer size that is too large will cause the source to move too far during averaging. Therefore, the buffer size used for the freely selected model path is reduced to 6,000 events, with re-averaging every 3,000 new events since the source moves much faster than the one following the path given by the DEM simulation. 

Compared to PEPT literature, the buffer sizes appear to be quite large (a typical sample size used to calculate a single PEPT position is of the order of 100). However, as introduced in the introduction, most PEPT detector systems used are based on crystal scintillators and a large number of readout channels (high granularity). Thus, the photopeak can be used to reduce background, and the high granularity leads to a very good spatial resolution. In order to achieve an equivalent resolution with lower granularity and missing photopeak, it is necessary to average over significantly more events.

All in all, the resolution of the tracer sphere position no longer corresponds to the spatial resolution of the detector system but has improved significantly.

\section{Results} \label{Sec:results}
\hfill\\[-0.31cm]
\noindent The objective of this simulation study was to extend the simulation and reconstruction of particles in granular assemblies to a moving source and to improve the accuracy of the reconstruction algorithm used. To this end, the spatial resolution (which is derived from the standard deviation of a Gaussian fit throughout this work) for a stationary source was first investigated at various locations in the detector system (\cref{Sec:boundaryEffects}). This was followed by the investigation of a source moving along a freely selected trajectory within the assembly (\cref{Sec:modelPath}) and a path given by a DEM simulation (\cref{Sec:demPath}). Finally, the model path was examined using the same buffer size as for the DEM path (\cref{Sec:modelPath_demBuffer}).

\subsection{Stationary source and boundary effects} \label{Sec:boundaryEffects}
\noindent To investigate the tracer particle position dependency of the reconstruction, the source was placed at different positions within the GGS. Each position was measured over a period of 1 second, corresponding to a total of $10^7$ simulated decays. Overall, it was found that the efficiency and resolution were nearly constant throughout the detector system and did not change significantly when the edges of the scintillator bars (i.e., the corners of the detector) were reached. This is due to the fact that the GGS is so small that the source is still far enough away from the edges of the detector and never touches them (compare \cref{Fig.detectorWithGGS}). Furthermore, decays that lie slightly outside the detector are also allowed for reconstruction, which significantly increases efficiency.

\begin{figure*}
\begin{subfigure}[b]{0.49\textwidth}
    \includegraphics[width=\textwidth]{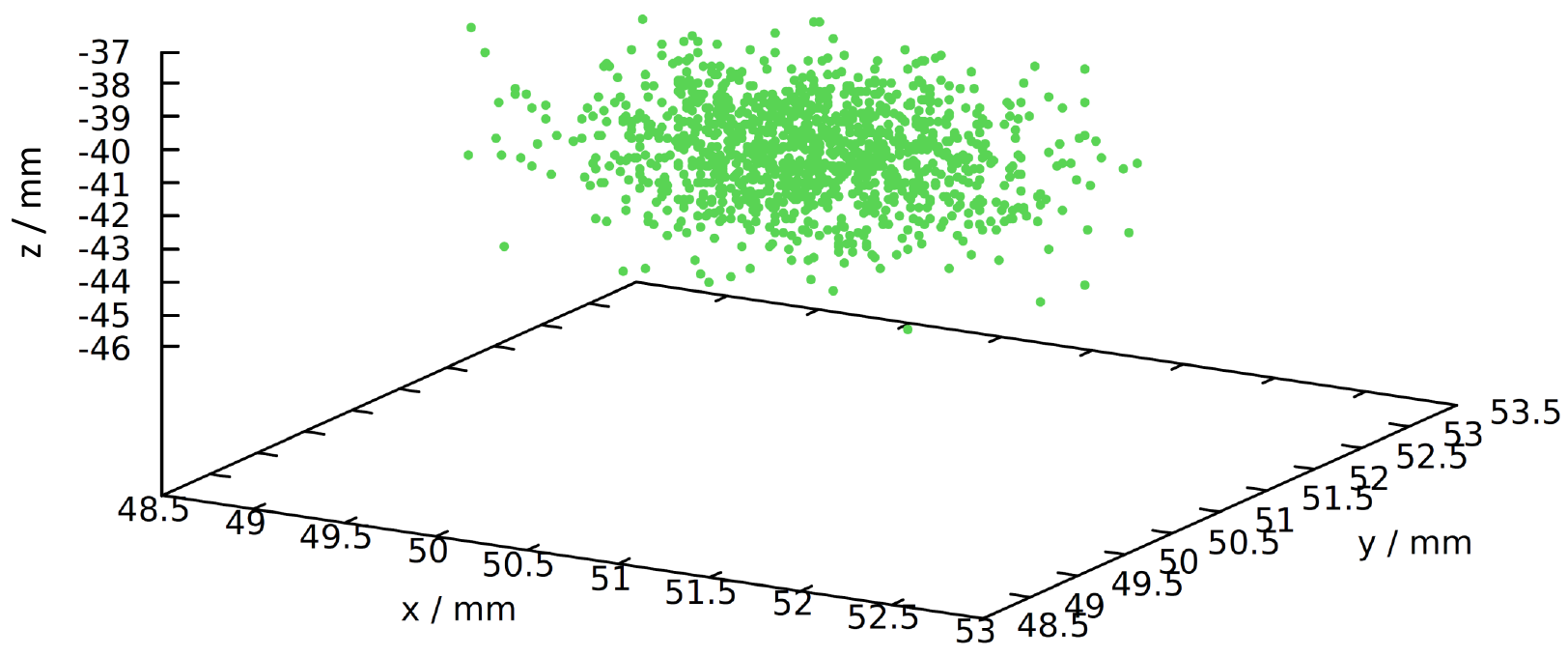}
    \caption{3D plot of the reconstructed events.}
    \label{Fig.stationary_small_3D}
\end{subfigure} 
\begin{subfigure}[b]{0.49\textwidth}
	\includegraphics[width=\textwidth]{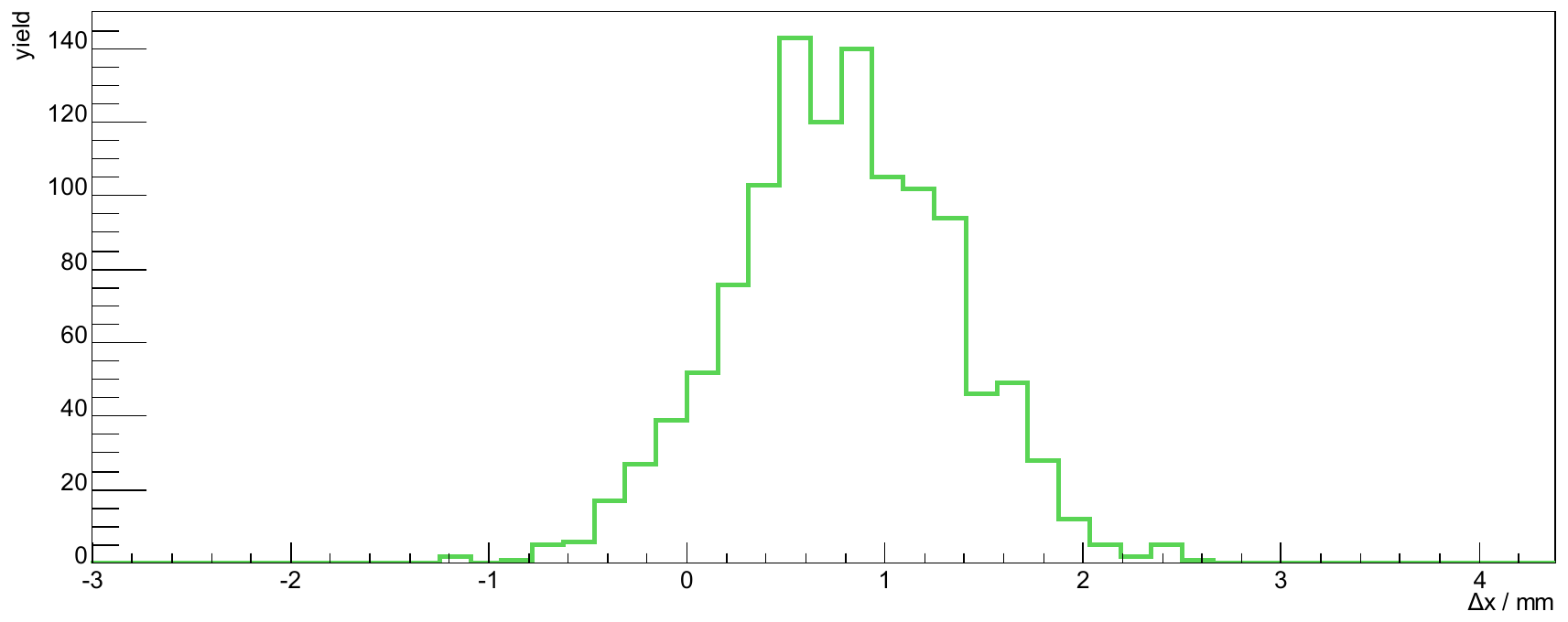}
    \caption{1D projection of the reconstructed events in x-direction.}
    \label{Fig.stationary_small_x}
\end{subfigure} 
\caption{Spatial distribution of the reconstructed events produced by a stationary source placed at (50\,|\,50\,|\,-40)\,cm after averaging and iterating. (a) shows the reconstructed events from a 3D perspective, and (b) its projection onto the x-axis.}
\label{Fig.stationarySource_small}
\end{figure*}

\cref{Fig.stationarySource} shows the three-dimensional distribution (\cref{Fig.stationary3D}) of the individual reconstructed tracer sphere positions without any averaging or iterating of a source that has been placed at (50\,|\,50\,|\,-40)\,cm within the GGS volume, as well as exemplarily its projection in the x-direction (\cref{Fig.stationaryX}). A very present background is visible, and the spatial resolution reads 104.3\,mm in the x-, 103.5\,mm in the y-, and 124.8\,mm in the z-direction. This is a very poor resolution since the size of the largest tracer sphere is much smaller than the best resolution obtained this way (103.5\,mm $\gg$ 20\,mm).
To improve this resolution, the simulated data were now averaged and iterated (as described in \cref{Sec:conditionsAndTracking}), resulting in a spatial resolution of 0.6\,mm in x and y, and 1.1\,mm in z. This corresponds to an improvement by a factor of about 170 in the x- and y-directions and a factor of 110 in the z-direction. The associated three-dimensional distribution and its projection on the x-axis can be found in \cref{Fig.stationarySource_small}.

\subsection{Model path} \label{Sec:modelPath}
\noindent To extend the simulation and reconstruction to a moving source and to investigate the behavior of the reconstruction algorithm at the highest possible velocities, a model curve was created on which the source moves circularly (r~=~10\,cm) at 17\,cm/s in the xy-plane and linearly between -6 and 2\,cm at 7\,cm/s in the z-direction. On average, these velocities approximate the maximum tracer sphere velocities within the GGS derived from corresponding DEM simulations. As mentioned in \cref{Sec:conditionsAndTracking}, the buffer size was set down to 6,000 events instead of 36,000, with re-averaging every 3,000 events to achieve good resolution. The measurement time of the whole curve was 4\,s.

\begin{figure}
 	\centering
    \includegraphics[width=0.48\textwidth]{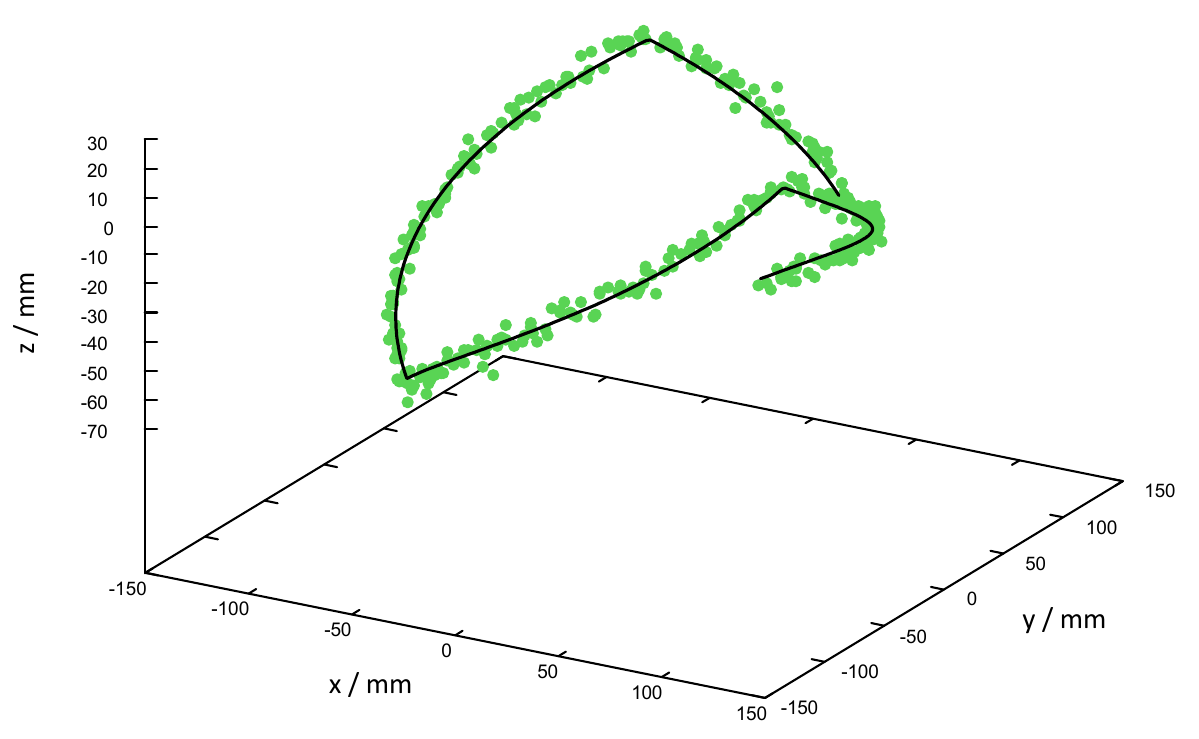}
    \caption{Simulated (black line) and reconstructed (green dots) model paths in a 3D representation.}
    \hfill \\[-0.32cm]
\label{Fig.modelPath3d}
\end{figure}

\begin{figure*}
\begin{subfigure}{\textwidth}
 	\centering
    \includegraphics[width=0.85\textwidth]{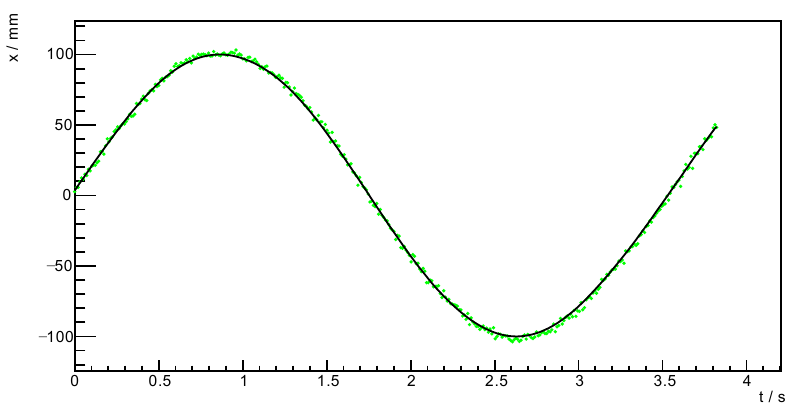}
    \label{Fig.modelx}
\end{subfigure} %
\hfill \\[-0.19cm]
\begin{subfigure}{\textwidth}
\centering
	\includegraphics[width=0.85\textwidth]{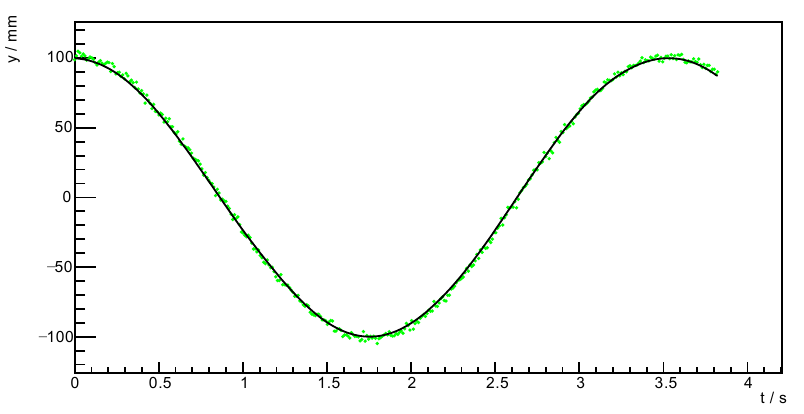}
    \label{Fig.modely}
\end{subfigure} %
\hfill \\[-0.19cm]
\begin{subfigure}{\textwidth}
\centering
	\includegraphics[width=0.85\textwidth]{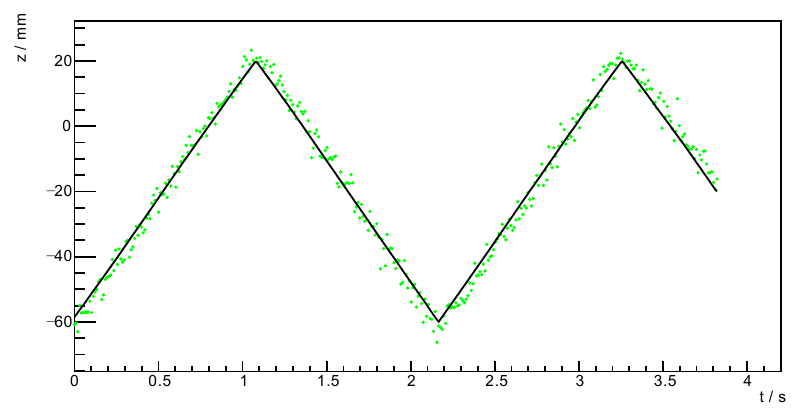}
    \label{Fig.modelz}
\end{subfigure} %
\hfill \\[-0.3cm]
\caption{Simulated (black line) and reconstructed (green dots) model paths.}
\label{Fig.modelPath}
\end{figure*}

A three-dimensional view of the model curve is shown in \cref{Fig.modelPath3d}, with the reconstructed tracer sphere locations shown in green and the actual simulated path as a black trajectory. The corresponding one-dimensional projections on the x, y, and z axes are shown in \cref{Fig.modelPath}. Both figures show that the reconstructed events cover the simulated trajectory very well. However, to estimate this more accurately, the distance of the reconstructed points from the simulated curve was determined, resulting in a spatial resolution of about 2.1\,mm in the x-, 2.2\,mm in the y-, and 3.0\,mm in the z-direction. Of course, the three-dimensional distance between the reconstructed tracer sphere positions and the simulated path can also be determined by:
\begin{align}
\Delta r = \sqrt{(\Delta x)^2 + (\Delta y)^2 + (\Delta z)^2}
\end{align}
which leads to an average deviation of about (4.0~$\pm$~1.6)\,mm. Thus, despite selecting the upper limits of the tracer particle velocities occurring in the grate system, all the spatial resolutions obtained are smaller than the size of the smallest spheres used: 3\,mm~<~10\,mm.

\subsection{Particle motion given by a DEM simulation} \label{Sec:demPath}
\noindent For the reconstruction of the DEM path, the position of the tracer sphere for each time step is taken from the DEM data file and inserted into the Monte Carlo simulations. Thus, only the decay location is moved instead of the whole volume. The measurement time for the entire path was 200\,s, although it was divided into fifty jobs at 4\,s. The DEM simulations have been carried out by an in-house code. Further details concerning the DEM code can be found in \ref{App:A}.

\begin{figure*}[H!]
\begin{subfigure}[b]{\textwidth}
 	\centering
    \includegraphics[width=0.99\textwidth]{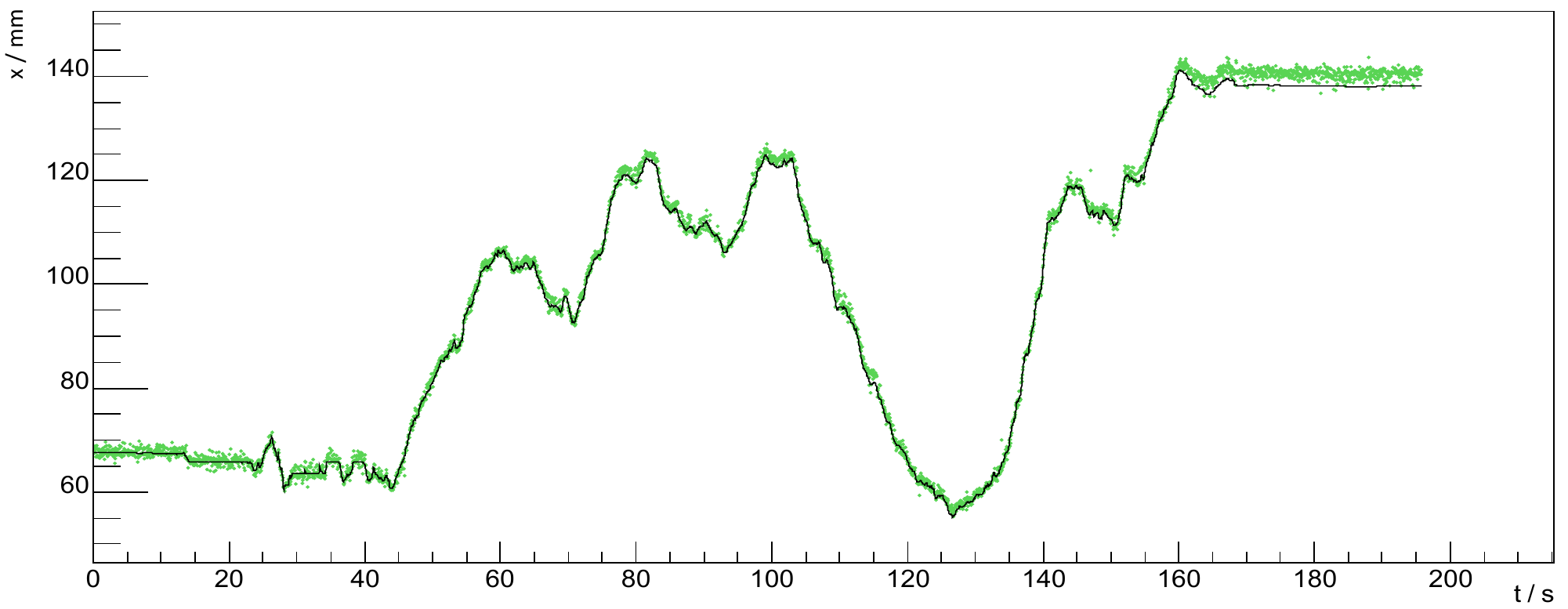}
    \label{demx}
\end{subfigure} %
\hfill \\[-0.1cm]
\begin{subfigure}[b]{\textwidth}
	\includegraphics[width=0.99\textwidth]{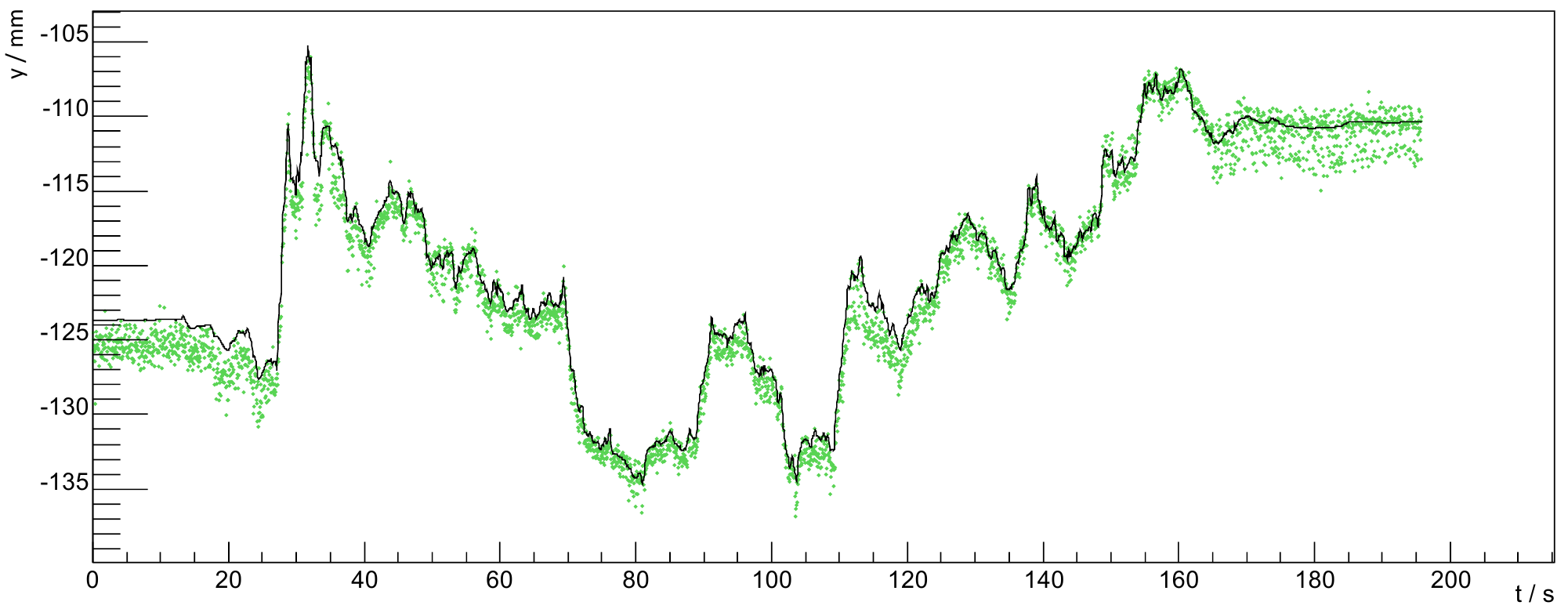}
    \label{demy}
\end{subfigure} %
\hfill \\[-0.1cm]
\begin{subfigure}[b]{\textwidth}
	\includegraphics[width=0.99\textwidth]{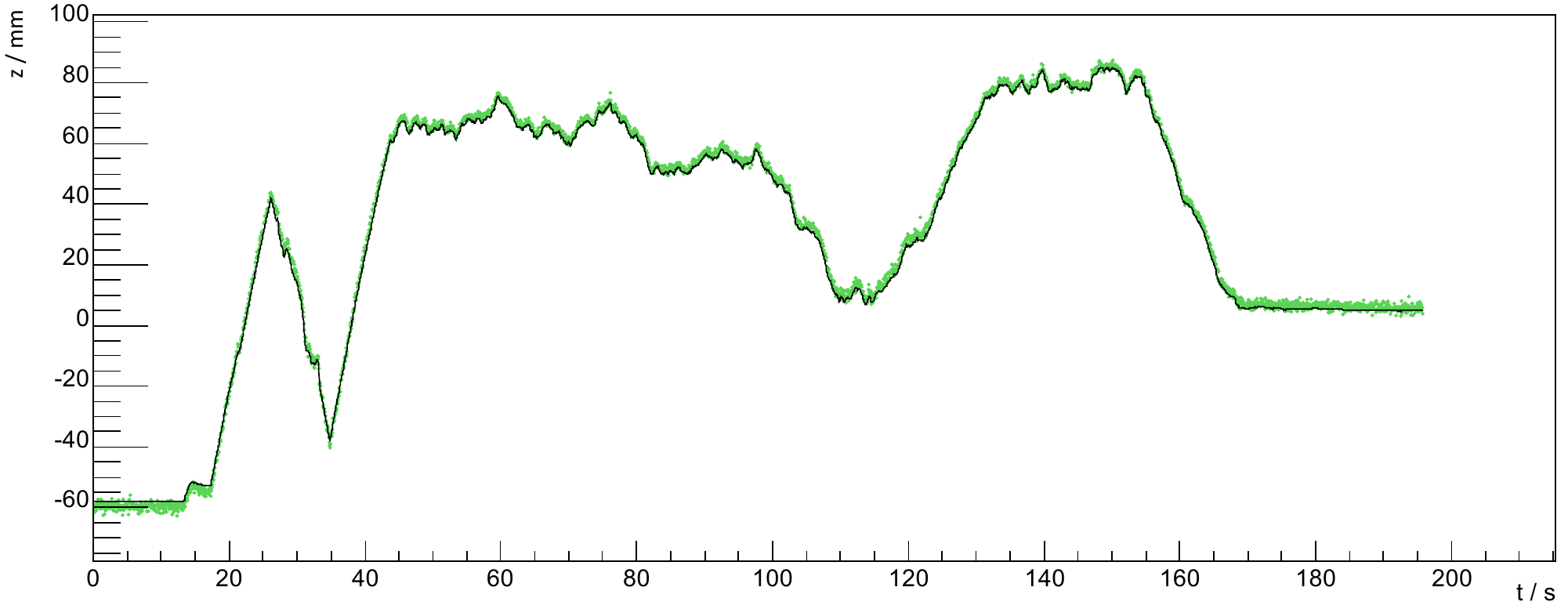}
    \label{demz}
\end{subfigure} %
\hfill \\[-0.3cm]
\caption{Simulated (black line) and reconstructed (green dots) paths of the tracer sphere based on DEM simulations.}
\label{Fig.DEMpath}
\end{figure*}

As can be seen in \cref{Fig.DEMpath}, the reconstructed tracer sphere positions cover the simulated data quite well, so a spatial resolution of 1.1\,mm in the x-, 1.0\,mm in the y-, and 1.3\,mm in the z-direction could be determined. The average three-dimensional distance between the reconstructed and simulated positions resulted in (2.3~$\pm$~0.9)\,mm. This is a very good result, as the value of the resolution is much smaller than the size of the smallest spheres used (1.3\,mm $\ll$ 10\,mm). Since the investigated path was given by a DEM simulation, it is a very realistic form of particle movement within the GGS, for which this work predicts a very good and satisfactory spatial resolution.

However, despite the good spatial resolution, \cref{Fig.DEMpath} shows a small systematic effect (the reconstructed points do not seem to be evenly distributed around the simulated data). Unfortunately, the cause of this shift has not been found yet, but it will be further investigated.

\subsection{Model path with same buffer size as in DEM simulation} \label{Sec:modelPath_demBuffer}
\begin{figure*}
\begin{subfigure}[b]{0.46\textwidth}
    \includegraphics[width=\textwidth]{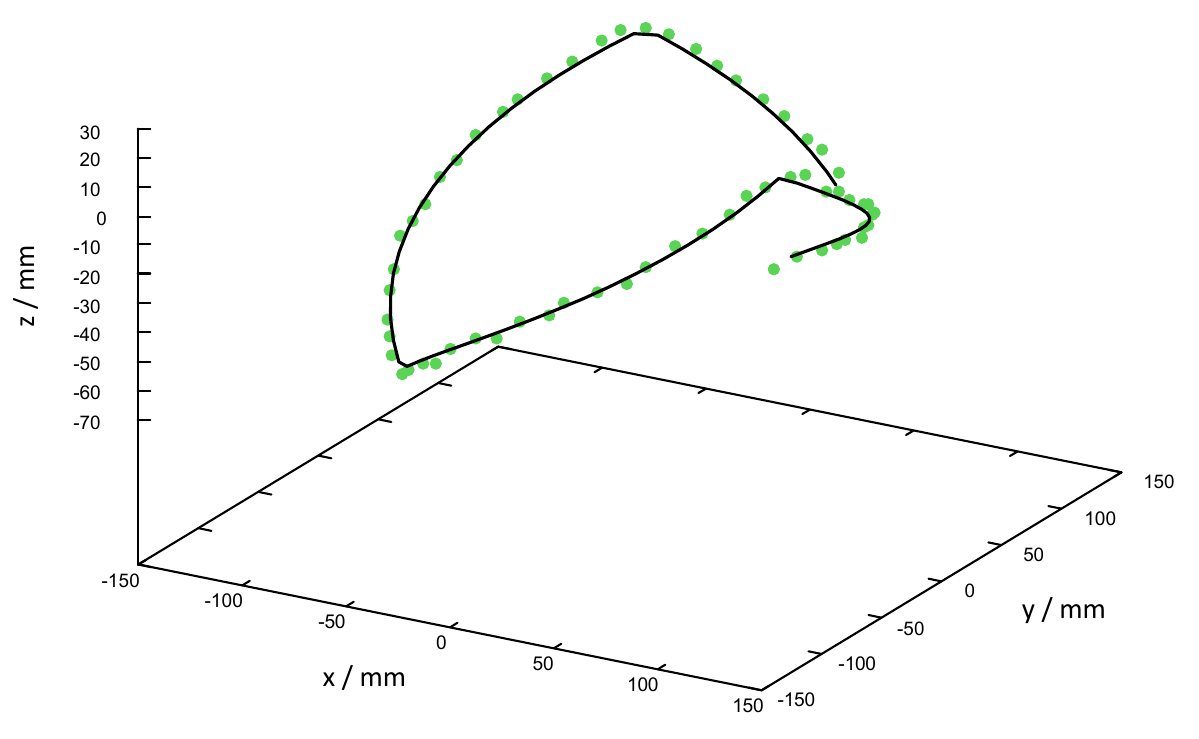}
    \caption{3D representation of the model curve.}
    \label{Fig.modelPath3d_demBuffer}
\end{subfigure} 
\begin{subfigure}[b]{0.53\textwidth}
	\includegraphics[width=\textwidth]{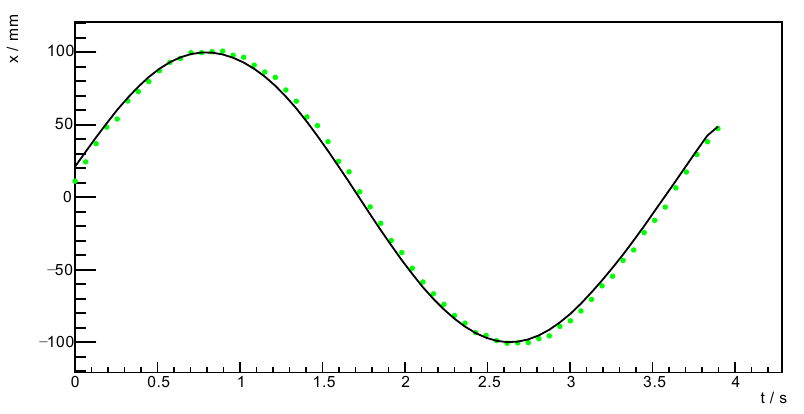}
    \caption{1D projection of the model curve in x.}
    \label{Fig.modelPathX_demBuffer}
\end{subfigure} 
\caption{Illustration of the reconstructed model path using the DEM buffer size (36,000 events). (a) shows a 3D perspective and (b) a projection of the path onto the x-axis. The reconstructed positions are shown as green dots, and the simulated path is shown as a continuous black line.}
\label{Fig.modelPath_demBuffer}
\end{figure*}

\noindent To illustrate the influence of the buffer size on the reconstruction of the tracer source and the resolution of the individual reconstructed tracer particle positions, the model curve was again examined, but this time with a significantly larger buffer size. This was done using exactly the same reconstruction as in \cref{Sec:modelPath} but this time with an enlarged buffer size of 36,000 events instead of the previously used 6,000 events. As can be seen in \cref{Fig.modelPath_demBuffer}, the density of reconstructed tracer sphere positions decreases significantly, which means that significantly fewer tracer sphere positions could be reconstructed in the same time interval, and the reconstructed positions are significantly farther apart from each other (i.e., the distances between each reconstructed data point are significantly larger than with a matched buffer size (\cref{Fig.modelPath3d})). However, this was to be expected, since using the same number of simulated events (again, 4\,s of measurement time) while averaging over 36,000 events every 18,000 new events instead of averaging over 6,000 every 3,000 new events results in at least 6 times fewer data points. As a consequence of the reduced statistics, the significance of the deviations is also reduced. Thus, the determination of the spatial resolution is not as meaningful as in all the other cases mentioned above. Nevertheless, the resolution determined with this data should be briefly presented for the sake of completeness. With only 62 events, the spatial resolution of the x-projection resulted in 4.2\,mm, the resolution of the y-projection in 4.0\,mm, the resolution of the z-projection in 3.3\,mm, and the average three-dimensional deviation in (6.5~$\pm$~1.5)\,mm.

Besides the lower statistics, there is also a small time shift. This is shown by a small but nevertheless uniform offset of the data points to the right (\cref{Fig.modelPathX_demBuffer}). A reasonable explanation for the time shift could be that the source moves significantly during the time that the 36,000 (or 18,000) events are collected for averaging.

Overall, it is evident that even at higher velocities/larger buffers, reconstruction of the tracer sphere positions is still possible, but the resolution and significance of the results are significantly reduced (three-dimensional deviation of about 6.5\,mm compared to 4.0\,mm with more reconstructed events/smaller buffer size). Therefore, attention should always be paid to ensure an appropriate adaptation of the reconstruction to the scenario being reconstructed.

\subsection{Comparison with data from previous paper} \label{Sec:comparison}
\noindent This section is intended to provide a comparison to the data obtained in the previous work (\cite{oppotsch2023simulation}), in which the reconstruction algorithm was tested in three different geometries with a resting tracer sphere. This resulted in a spatial resolution of about 1.2 - 1.5\,mm for the detector only filled with air, a spatial resolution of a resting source placed inside the generic grate system of about 4.7 - 5.9\,mm, and a spatial resolution of the detector system completely filled with particles of about 9.7 - 12.7\,mm.

After scaling down the detector, as well as adjusting and improving the reconstruction algorithm, it was possible to determine a stationary source within the GGS with a spatial resolution of 0.6\,mm in the x- and y-directions and 1.1\,mm in the z-direction. Thus, the resolution was improved by about 5.2\,mm in the x- and y-directions and 3.6\,mm in the z-direction. Even compared to the results obtained with the detector system under ideal conditions (filled with air only so that no scattering or absorption occurs), there is an improvement in spatial resolution of about 0.6\,mm in the x- and y-directions and an improvement of 0.4\,mm in the z-direction. 

Considering the case of a moving source (whether on the model curve or the DEM path), once more, a significantly better resolution within the GGS than before the above-mentioned adjustments is seen. Looking specifically at the particle motion resulting from a DEM simulation, the spatial resolution again exceeds the results obtained for the empty detector system (improvement of about 0.2\,mm) and is thus better than the previous resolution in the GGS by about 4.8\,mm, 4.7\,mm, and 3.4\,mm, respectively.

All spatial resolutions and three-dimensional deviations determined in this work can also be looked up in \cref{Tab:spatialResolutions} and \cref{Tab:3dDeviations}.

\begin{table}
	\centering
	\caption{Overview of all spatial resolutions.} 
	\begin{tabular}{l l}
	\toprule[1pt]
	Scenario						  & Spatial resolution (mm) \\
	\toprule[1pt]
	Stationary source 				  & $\Delta$x = 0.559 \\
	                                  & $\Delta$y = 0.566 \\
	                                  & $\Delta$z = 1.145 \\
	\hline	\\[-8pt]
	Model path (buffer size = 6,000)  & $\Delta$x = 2.128 \\
									  & $\Delta$y = 2.232 \\	
									  & $\Delta$z = 3.036 \\
	\hline	\\[-8pt]
	DEM path						  & $\Delta$x = 1.095 \\
	                                  & $\Delta$y = 0.995 \\
	                                  & $\Delta$z = 1.304 \\
	\hline	\\[-8pt]
	Model path (buffer size = 36,000) & $\Delta$x = 4.215 \\
	                                  & $\Delta$y = 4.035 \\
	                                  & $\Delta$z = 3.256 \\[1.5pt]
	\toprule[1pt]
	\end{tabular}
\hfill \\[-0.212cm]
\label{Tab:spatialResolutions}
\end{table}

\section{Conclusion and outlook} \label{Sec:outlook}
\noindent This work is the second part of a simulation study investigating the performance of an envisaged cost-effective PET-like detector system capable of tracking tracer particle movement in dense and optically opaque granular assemblies. Following on from the work done in \cite{oppotsch2023simulation} (and therefore referred to as the previous studies for the rest of this section), which investigated the performance of the detector system for a stationary source within the detector system filled with air only (referred to as the empty detector), the generic grate system and the detector system completely filled with particles, the goal of this work was to improve the existing reconstruction software (by extending it to oversampling (calculating a moving average for the position determination of the tracer particle) and iteration) and extend both the reconstruction and simulation software to a tracer sphere moving along a selected model path and a path given by a DEM simulation. Moreover, the PET-like detector system was scaled down to a total of 88 cost-effective plastic scintillator bars to better fit the size of the generic grate system and to further reduce cost. 

\begin{table}
	\centering
	\caption{Three-dimensional deviations for all scenarios.} 
	\begin{tabular}{l l}
	\toprule[1pt]
	Scenario						  & Deviation $\Delta$r (mm) \\
	\toprule[1pt]
	Stationary source 				  & 2.1 $\pm$ 0.8 \\
	Model path (buffer size = 6,000)  & 4.0 $\pm$ 1.6 \\
	DEM path						  & 2.3 $\pm$ 0.9 \\
	Model path (buffer size = 36,000) & 6.5 $\pm$ 1.5 \\[1.5pt]
	\toprule[1pt]
	\end{tabular}
\label{Tab:3dDeviations}
\end{table}

Thus, for a stationary source within the GGS, spatial resolutions of 0.6\,mm in the x- and y-direction and 1.1\,mm in the z-direction, as well as an average three-dimensional deviation of the reconstructed tracer sphere positions from the simulated position of (2.1~$\pm$~0.8)\,mm were determined. Furthermore, it could be shown that the iteration leads to an improvement of the resolution by a factor of about 170 in x and y and by a factor of 110 in the z-direction. By measuring the position dependency of the reconstruction, it was found that the efficiency is nearly constant throughout the detector system (due to the small size of the generic grate system), so boundary effects can be neglected. 

To investigate the spatial resolution of a source moving at maximum velocity within the GGS, a model curve was developed. On this curve, the tracer sphere performed a circular motion (radius 10\,cm) with a velocity of 17\,cm/s in the x- and y-direction and a linear motion (between ${\text{-6}}$ and 2\,cm) with a velocity of 7\,cm/s in the z-direction. The reconstruction was executed with two different buffer sizes: A smaller buffer of 6,000 events for each averaging, where every new averaging was done after 3,000 newly collected events, and a larger buffer size (36,000 events per buffer and a re-averaging every 18,000 new events) chosen to meet the DEM path conditions. For these two cases, it became clear that the buffer size should not be too large for higher velocities. Otherwise, the gaps between the individual data points become relatively large. In addition, the fast motion of the sphere during data acquisition results in a small time shift for each buffer (visible by a small but uniform offset of the reconstructed tracer sphere positions to the simulated positions in the projections onto the three axes). However, even if a relatively good spatial resolution has been achieved, it is not very meaningful due to the paucity of statistics. With the smaller buffer, these phenomena did not occur, so a good resolution of 2.1\,mm in x, 2.2\,mm in y, and 3.0\,mm in z could be achieved. The three-dimensional deviation was (4.0~$\pm$~1.6)\,mm on average.

Finally, a path taken from a DEM simulation was examined as a trajectory for the source. Here, an excellent spatial resolution of 1.1\,mm in x, 1.0\,mm in y, and 1.3\,mm in z, with a three-dimensional deviation of (2.3~$\pm$~0.9)\,mm was obtained. Thus, there is an improvement of about 0.2\,mm compared to the previous studies within the empty detector (representing the best results possible since the gamma rays are not scattered or absorbed). With respect to the previous studies within the generic grate system, the spatial resolution could be improved by 4.8\,mm in the x-direction, 4.7\,mm in the y-direction, and 3.4\,mm in the z-direction. These are pretty good results since all spatial resolutions are better than the smallest size of the spheres used (1.3\,mm~$\ll$~10\,mm).

Next up is the extension of the existing reconstruction algorithm to clustering to ensure better cross-checking and a straightforward extension to multiple sources. The most recent and promising methods are introduced in the review article by \cite{PEPTReview}. However, extension to multiple sources can also be achieved by extending the likelihood fit, which was introduced and tested in \cite{oppotsch2023simulation}. 

Another pending issue is the investigation of the influence of a $^{22}$Na source on the reconstruction. This is necessary because such a source will be used in the upcoming measurements to generate the 511\,keV gamma-ray pairs for the reconstruction, but the simulations have so far been performed with 511\,keV gamma rays only. During the decay of the $^{22}$Na source, gamma rays with an energy of 1275\,keV are emitted as well. Hence, the influence of these additional gamma rays on the tracer particle reconstruction has to be investigated.

\hfill \\[-0.5cm]
\section*{Declaration of competing interest}
\noindent The authors declare that they have no known competing financial interests or personal relationships that could have appeared to influence the work reported in this paper.

\hfill \\[-0.5cm]
\section*{Acknowledgments}
\noindent This work has been funded by the Deutsche Forschungsgemeinschaft (DFG, German Research Foundation) -- Project-ID 422037413 -- TRR 287.

\appendix
\renewcommand{\thesection}{Appendix A} 
\hfill \\[-0.5cm]
\section{DEM simulations of the generic grate system}\label{App:A}
\noindent The discrete element method (DEM) simulations were performed with an in-house code. The equations for translational and rotational motion are given by
\begin{align}
& m_i \frac{d^2 \vec{x}_i}{dt^2} = \sum\limits_{j=1}^{N} \vec{F}_{ij} + m_i \vec{g} \\
& \Theta_i \frac{d^2 \vec{\phi}_i}{dt^2} = \sum\limits_{j=1}^{N} \vec{M}_{ij} = \sum\limits_{j=1}^{N} \left(\vec{r}_i \times \vec{F}_{ij} + \vec{M}_j^r\right)
\end{align}
where $m_i$ is the mass of the particle, and its moment of inertia is $\Theta_i$. The linear acceleration is given by $d^2\vec{x}_i / dt^2$ and the angular acceleration of the particle is $d^2\vec{\phi}_i / dt^2$. $\vec{F_{ij}}$ is the external force and $\vec{M}_{ij}$ the external momentum induced by other particles or walls. $\vec{M}^r_j$ is the rolling friction torque. The distance from the center of gravity to the contact point of particle/particle or particle/wall is represented by $\vec{r}_i$. The equations are integrated numerically by an Euler-Cromer algorithm. To determine the contact forces, a linear spring-dashpot model is used. The parameters used for the simulation of the generic grate system can be found in \cite{PaperNikoline} and are listed in \cref{Tab:DemParameters}. 

\begin{table*}
\centering
\caption{Collision parameters for DEM simulations.} 
	\begin{tabular}{p{5cm} p{3.5cm} p{3.5cm} p{3.5cm}}
	\toprule[1pt]
	  & Sphere on sphere & Sphere on bar & Sphere on wall \\
	\toprule[1pt]
	Coefficient of restitution $e^n$  		& 0.926  & 0.870 & 0.838 \\
	Coefficient of static friction $\mu_c$   & 0.242 & 0.178  & 0.191 \\
	Coefficient of rolling friction $\mu_r$  & 		 & 		  & 		 \\
	\hline	\\[-8pt]
	~~~~~~~Sphere 10 mm  & 5.257 $\times$ $10^{-5}$ & 4.180 $\times$ $10^{-5}$ & 2.973 $\times$ $10^{-5}$ \\
	~~~~~~~Sphere 15 mm  & 4.702 $\times$ $10^{-5}$ & 4.041 $\times$ $10^{-5}$ & 2.874 $\times$ $10^{-5}$ \\	
	~~~~~~~Sphere 20 mm  & 4.146 $\times$ $10^{-5}$ & 3.901 $\times$ $10^{-5}$ & 2.775 $\times$ $10^{-5}$ \\[1.5pt]
	\toprule[1pt]
	\end{tabular}
\label{Tab:DemParameters}
\end{table*}

Note that the code has been extensively tested in the past to ensure correct particle mechanics. Over the past 15 years, examples comprise the simulation of vibrating conveyors \citep{simsek2008experimental}, silo discharge \citep{komossa2014transversal}, rotary drums \citep{hohner2015study}, or grate systems \citep{PaperNikoline,sudbrock2011discrete}. An overview of other applications of the code, including chemical reaction, is given in \cite{mahiques2023simulation}.

For the present study, an exemplary particle trajectory has been selected from the DEM simulations of the generic grate system shown in \cref{Fig.sketch}. The idea is here just to provide a typical trajectory of a potential tracer particle as an input to the Monte Carlo simulations to reconstruct particle position.

\bibliography{PaperBib}

\end{document}